# A quantum theory of thermodynamic relaxation

Roumen Tsekov
Department of Physical Chemistry, University of Sofia, 1164 Sofia, Bulgaria

A new approach to quantum Markov processes is developed and the corresponding Fokker-Planck equation is derived. The latter is examined to reproduce known results from classical and quantum physics. It was also applied to the phase-space description of a mechanical system thus leading to a new treatment of this problem different from the Wigner presentation. The equilibrium probability density obtained in the mixed coordinate-momentum space is a reasonable extension of the Gibbs canonical distribution.

Despite the great progress in contemporary quantum statistical physics [1], there are still problems in the applicability of the developed concept to complex systems such as in chemistry. As a rule, the theory of quantum relaxation is much less elaborated than that of the equilibrium. This fact is not surprising since the same situation holds in classical physics. In the latter, the most powerful theory of relaxation phenomena is due to the Markov approximation of processes in Nature. There are some attempts to develop a corresponding quantum idealization [2], e.g. Glauber-Sudarshan [3] and Husimi [4] presentations, but they are less universal. The most general approach to quantum dissipation consists in the projection of the Universe dynamics to those of a system and environment. This means that after integration along the environment variables of the rigorous von Neumann equation one can yield equations for the system evolution. Due to obvious mathematical complications it is only applied to coupled harmonic oscillators [5, 6]. A way to slide over the many-particle problems is to introduce dissipative Schrödinger equations [7-9]. In the literature, other methods are also reported which are based either on the stochastic treatment of the Schrödinger equation [10, 11] or on the Langevin-like description of quantum dynamics in the Heisenberg time-dependent operator formalism [12, 13].

The present paper aims to generalize our previous approach to the Markov quantum theory [14-16], which, suffering from a rigorous derivation, yields good results. Last but not the least advantage of our theory is its physical transparency and relative universality. To make the consideration plausible, let us start with a brief review of the quantum formalism. Suppose there is a mechanical system isolated from all the rest in the world (in fact, the only example we have is the Universe). According to quantum mechanics, the system wave function in a $a$-representation ($a$ can be either momentum or coordinate) obeys the Schrödinger equation $i\hbar \partial_t \psi_a = \hat{H}_a \psi_a$, where $\hat{H}_a$ is the system Hamiltonian operator. The solution of the Schrödinger equation is a set of eigenvalues and normalized eigenfunctions $\{E_n, \varphi_n\}$ of the system Hamiltonian. Therefore, if the energy of the system is known, one is able to determine the corresponding state and probability density. System energy, however, is not an experimentally accessible pa-rameter and, for this reason, the above program is not a working one. One can measure only energy difference, which is sufficient for description of the involved process only. For instance, the energy levels of a harmonic oscillator are equidistant and there is no experimental way to guess which state the system occupies at present.

The most popular system in chemistry is that exchanging energy with the surrounding medium at constant temperature $T$. It is a particular example for a system-environment division and the temperature is the basic characteristic of the environment. After Gibbs, there is no difficulty in obtaining the equilibrium probability density via the canonical expression

$$\rho_e(a,\beta,n) = Z^{-1} \exp(-\beta E_n)\varphi_n^2(a) \tag{1}$$

where $Z = \sum \exp(-\beta E_n)$, $\beta = (k_B T)^{-1}$ and $k_B$ is the Boltzmann constant. However, as mentioned before, there is no general description of the non-equilibrium dynamics of the isothermic system, yet. The rigorous way involves so many integrals of the exact von Neumann equation, which are out of the limits even for computers of our century. For this reason, the existing kinetic theories [1-3] are applicable only to some simple model systems with unrealistic coupling interaction between system and environment [17-19]. In the present study, a new model of irreversible dynamics in an isothermic quantum system is developed which is based on a linear non-equilibrium thermodynamic treatment of the problem. One can easily prove that distribution (1) is the solution of the following stationary Schrödinger equation $\hat{H}_a^\beta \psi_a^\beta = E\psi_a^\beta$ with a modified Hamiltonian operator $\hat{H}_a^\beta = \hat{H}_a + \partial_\beta + \partial_\beta^+$ (the superscript $^+$ indicates Hermitian conjugation). Here $\psi_a^\beta$ is the temperature dependent wave function and the probability density is defined as the square of the wave function. The modified Schrödinger equation can be rewritten in the form

$$\partial_a[(\psi_a^\beta)^{-1}\hat{H}_a\psi_a^\beta] - \partial_\beta(\beta\phi_a) = 0 \tag{2}$$

thus representing the static force balance between the system and its environment. The thermal force functional $\phi_a = -k_B T \partial_a \ln \rho_e$ plays a key role in the description of thermodynamic relaxation [2, 20, 21].

Let us consider now the irreversible dynamics of the isothermic system. The time evolution of the probability density should necessarily satisfy the continuity equation

$$\partial_t \rho + \partial_a(\rho v) = 0 \tag{3}$$

which follows from the von Neumann equation and can be adopted as definition of the velocity $v$ in the system probability space. For the sake of simplicity we shall restrict the further approach to the range of slow propagating processes, a necessary condition for a Markov idealization. Hence, in the frame of linear non-equilibrium thermodynamics [2], the following two equations hold

$$\phi_a = -k_B T \partial_a \ln \rho - L_a^{-1} v + o(v^2) \qquad \text{Re}\,[(\psi_a^\beta)^{-1} \hat{H}_a \psi_a^\beta] = \rho^{-1/2} \hat{H}_a \rho^{1/2} + o(v^2)$$

where $L_a$ is the matrix of kinetic coefficients. The second relation presumes no existence of imaginary terms in the system Hamiltonian and for this reason there is no term linear in $v$ there. Substituting these relations in (2) and integrating the result once over $\beta$, one obtains an expression for the velocity in the probability space

$$v = -L_a \partial_a F_a \qquad (4)$$

where the integral

$$F_a = \int_0^1 \rho(a, x\beta, t)^{-1/2} \hat{H}_a^{x\beta} \rho(a, x\beta, t)^{1/2} dx$$

represents the characteristic thermodynamic functional of the system free energy [2] and possesses a minimum at the equilibrium state. In the latter case we have $\hat{H}_a^\beta \rho_e^{1/2} = E \rho_e^{1/2}$, which provides as equilibrium solution the Gibbs distribution (1). It should be noted that the force balance (2) is employed without addition of any inertial term because of the restriction to slow variables. Now, introducing (4) in the continuity equation (3) yields an equation governing the evolution of the probability density

$$\partial_t \rho = \partial_a (\rho L_a \partial_a F_a) \qquad (5)$$

This Fokker-Planck-like equation [9] is the basic result of the present Markov theory.

To demonstrate the correctness of (5), we can compare its predictions with some known results. If one is interested in the evolution of the probability density in the momentum space, $a = p$, the corresponding kinetic coefficient matrix $L_p$ is equal [2,3] to that of the friction coefficients $B$. For a system of a free ideal gas the Hamiltonian is given by $\hat{H}_p = pM^{-1}p/2$, where $M$ is the mass matrix. In this case (5) transforms to

$$\partial_t \rho = \partial_p (\rho B M^{-1} p + k_B T B \partial_p \rho) \qquad (6)$$

This is the well-known Fokker-Planck equation in the momentum space [3]. Note that the generalized Fokker-Planck equation for the momentum space always reduces to (6) in the classical limit. The second interesting case is the evolution in the configuration space, $a = q$, where $L_q = B^{-1}$. In the classical limit $\hbar \to 0$ the corresponding equation (5) reduces to equation in the position space

$$\partial_t \rho = \partial_q (\rho B^{-1} \partial_q U + k_B T B^{-1} \partial_q \rho) \tag{7}$$

where $U$ is the potential energy of the system. Equation (7) is known in the literature after the name of Smoluchowski [2].

To illustrate the correctness of (5) in the quantum case, we shall briefly present here results for a harmonic oscillator with Hamiltonian operator $\hat{H}_q = -(\hbar^2/2m)\partial_q^2 + m\omega^2 q^2/2 - fq$, where $f$ is an external force. The solution of (5) for this case is a normal distribution density [14]. At large times, stationary state is established with oscillations' dispersion $\sigma^2 = (\hbar/2m\omega)\coth z$ with $z = \beta\hbar\omega/2$. Then, the evolution of the mean value $\bar{q}$ obeys the following equation [14]

$$\partial_z(zb\partial_t\bar{q}) + m\omega^2(\bar{q} + \tanh z\, \partial_z\bar{q}) = f \tag{8}$$

In the classical limit, this equation acquires the well-known form $b\partial_t\bar{q} + m\omega^2\bar{q} = f$. In the opposite case of low temperatures, according to the transition state theory, the friction coefficient $b$ is inversely proportional to the oscillator partition function, $b^{-1} \sim z/\sinh z$. Using this relation and a new variable $y = \bar{q} + \tanh z\, \partial_z\bar{q}$, equation (8) can be rewritten as $bz\coth z\, \partial_t y + m\omega^2 y = f$. As is seen, the friction term here differs from that in the classical case; the expression $bz\coth z$ for the friction coefficient is known in the literature as the quantum fluctuation-dissipation relation [12].

Finally, we are tempted to examine (5) in the system phase-space $a = (p, q)$. Apart from the Wigner function there are many other distributions proposed to describe the quantum systems in their phase-space [4]. A natural extension for the kinetic coefficient matrix from the previous two cases is a super-matrix with elements $L_{pp} = L_{qq}^{-1} = B$ and $L_{pq} = -L_{qp} = I$, where $I$ is the unit tensor. This presentation is consistent with the Onsager symmetry rule [2] and the corresponding equation (5) has the form

$$\partial_t\rho + \partial_q(\rho\partial_p F_p) - \partial_p(\rho\partial_q F_q) = \partial_p(\rho B \partial_p F_p) + \partial_q(\rho B^{-1}\partial_q F_q) \tag{9}$$

Since $F_a$ is invariant of the sign of $\rho$, the above equation provides only non-negative solutions and thus eliminates the basic shortcoming of the Wigner function method. In contrast to other methods [17-19, 22], (9) is an integrodifferential equation of finite order and in this way it is a simpler mathematical problem. To find the general solution of (9) is not trivial. It shows, however, that the equilibrium distribution obeys simultaneously the following system of two equations $\hat{H}_q^\beta \rho_e^{1/2} = E\rho_e^{1/2}$ and $\hat{H}_p^\beta \rho_e^{1/2} = E\rho_e^{1/2}$. Hence, the equilibrium probability density in the system phase-space is

$$\rho_e(q, p, \beta, n) = Z^{-1} \exp(-\beta E_n)\varphi_n^2(q)\varphi_n^2(p) \tag{10}$$

which seems to be a reasonable extension of the canonical Gibbs distribution (1). This equation demonstrates statistical independence of the momentum and coordinate in the frames of a given quantum state. Using (10) one can calculate the average value of any function of $q$ and $p$. The classical analogue of (9) yielded in the limit $\hbar \to 0$ is

$$\partial_t \rho + pM^{-1}\partial_q \rho - \partial_q U \partial_p \rho = \partial_p (\rho BM^{-1} p + k_B TB \partial_p \rho) + \partial_q (\rho B^{-1} \partial_q U + k_B TB^{-1} \partial_q \rho) \qquad (11)$$

The equilibrium solution of this equation is the Maxwell-Boltzmann distribution. In the large friction limit the last term omits and the result is known in the literature as the Klein-Kramers equation. Equation (11) has been first proposed by Ilya Prigogine.